\documentclass[epj,final]{svjour}
\usepackage{graphics}
\begin{document}
\title{Experimental Tests of Non-Perturbative Pion Wave Functions} 
\author   {Daniel Ashery \inst{1}
                   \and
	   Hans-Christian Pauli \inst{2} }
\institute{School of Physics and Astronomy,
	   Raymond and Beverly Sackler Faculty of Exact Sciences,
	   Tel Aviv University, 
	   Israel,
	   \email{ashery@tauphy.tau.ac.il}
	           \and 
           Max-Planck-Institut f\"ur Kernphysik,
           D-69029 Heidelberg,
	   \email{pauli@mpi-hd.mpg.de}}
\offprints{H.C. Pauli, 
	   Postfach 10 39 80,
	   D-69029 Heidelberg}
\mail     {D. Ashery, 
	   School of Physics and Astronomy,
           Raymond and Beverly Sackler Faculty of Exact Sciences, 
           Tel Aviv University,
	   Tel Aviv,
	   Israel}
\date{14 January 2003}
\abstract{We use the transverse-momentum dependence of the cross section
for diffractive dissociation of high energy pions to two jets to study 
   some non-perturbative Light-Cone wave functions of the pion. 
   We compare the predictions for this distribution by Gaussian and  
   Coulomb wave functions as well as the wave function derived from 
   solution of the Light-Cone Hamiltonian in the Singlet Model. 
   We conclude that this experimentally measured information 
   provides a powerful tool for these studies. 
\PACS{%
       {11.15.Tk}  
  \and {12.38.Lg}  
  \and {12.39.-x}  
  \and {13.90.+i}{}
  } 
}
\maketitle
\section{\large \bf Light-Cone Wave Functions}
\label{sec:int} 

One of the most interesting subjects of particle and nuclear physics is
understanding the internal structure of hadrons. It bears directly on the 
fundamental interactions of quarks and gluons that create the hadronic 
bound state. It is also an essential ingredient in understanding the 
hadronic
strong, electromagnetic and weak interactions. A very powerful description
of the hadronic structure is obtained through the light-cone wave 
functions.
These are frame-independent and process-independent quantum-mechanical
descriptions at the amplitude level. 
They encode all possible quark and gluon
momentum, helicity and flavor correlations in the hadron. 
The light-cone wave functions are constructed from the 
QCD light-cone Hamiltonian \cite{BroHwaMa01}:
$H_{LC}^{QCD} = P^+P^- - P^2_{\perp}$, where 
$P^{\pm} = P^0 \pm P^z$. 
The wave function 
$\psi_h$ for a hadron $h$ with mass $M_h$ 
satisfies the relation: 
$H_{LC}^{QCD}\vert\psi_h\rangle = M_h^2 \vert\psi_h\rangle$.

The light-cone wave functions are expanded in terms of a complete basis of 
Fock states having increasing complexity. In this way the hadron 
presents itself as an ensemble of coherent states containing various
numbers of quark and gluon quanta \cite{Bro98}. For example, the negative 
pion has the Fock expansion:
\begin{eqnarray*}
   \vert\psi_{\pi^-}\rangle 
   &=& \sum _n \langle n\vert\pi^-\rangle \vert n\rangle \\
   &=&\psi^{(\Lambda)}_{d\bar{u}/\pi}(u_i,\vec{k}_{\perp i})
   \vert\bar{u}d\rangle +
   \psi^{(\Lambda)}_{d\bar{u}g/\pi}(u_i,\vec{k}_{\perp i})
   \vert\bar{u}dg\rangle  + \dots 
\end{eqnarray*}
representing expansion of the exact QCD eigenstate at scale $\Lambda$ 
in terms of non-interacting quarks and gluons.
They have longitudinal light-cone momentum fractions:
\begin{equation}
   u_i = \frac {k_i^+}{p^+} = \frac {k_i^0+k_i^z}{p^0+p^z}\;, \qquad
   \sum_{i=1}^n u_i =1
\;,\label{eq:xlc}\end{equation}
and relative transverse momenta
\begin{equation}
   \vec{k}_{\perp i} \;, \qquad
   \sum_{i=1}^n \vec{k}_{\perp i} = \vec 0_{\perp} 
\;.\label{eq:kplc}\end{equation}
The form of $\psi_{n/H}(u_i,\vec{k}_{\perp i})$ is invariant under
longitudinal and transverse boosts; i.e., the light-cone wave functions 
expressed in the relative coordinates $u_i$ and $k_{\perp i}$ are 
independent of the total momentum ($p^+$, $\vec{p}_{\perp}$) of the
hadron. The Fock states are off mass shell with masses of 
$M_n = \sum_{i=1}^n(k_{\perp i}^2+m_i^2)/u_i$
where $m_i$ are the quark (current) masses.
The first term in the expansion is referred to as the valence Fock state,
as it relates to the hadronic description in the constituent quark model.
The higher terms are related to the sea components of the hadronic 
structure.
It has been shown that once the valence Fock state is determined it is
possible to build the rest of the 
light cone wave function \cite{Mue94,AntBroDal97}.
This was done for the pion using Discretized Light-Cone Quantization 
(DLCQ) on transverse lattice \cite{bpp97,Dal00}. 

\begin{figure*}\sidecaption
   \resizebox{0.70\textwidth}{!}{
   \includegraphics{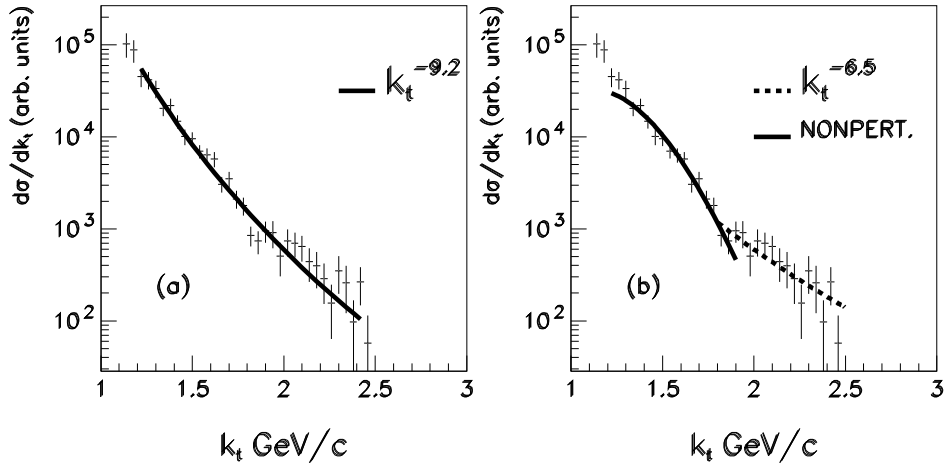}
}\caption{Comparison of the $k_t$ distribution of 
   acceptance corrected data with fits to
   cross section dependence (a) according to a power law, (b) based on a
   nonperturbative Gaussian wave function for low $k_t$ and a power
   law, as expected from perturbative calculations, for high $k_t$.
   Reproduced from Ref.\protect{\cite{e791}}.
}\label{split}
\end{figure*}

The hadronic distribution amplitude $\phi(u,Q^2)$ is the 
probability amplitude to find a quark and antiquark of the respective 
lowest-order Fock-state carrying  fractional momenta $u$ and $1-u$ 
\cite{leb80,BroLep81}. 
The pion distribution amplitude and the light-cone wave function of the
respective  Fock state $\psi$ are related through \cite{leb80,BroLep81}:
\begin{eqnarray}
   \phi_{q\bar{q}/\pi}(u,Q^2)  &\sim&  \int_0^{Q^2} \psi_{q\bar{q}/\pi}
   (u,\tilde{k_{\perp}}) d^2\tilde{k_{\perp}}
\;,\label{phihad}\\
  Q^2  &=&  \frac{k_{\perp}^2}{u(1 - u)}
\;.\end{eqnarray}
Two functions have been proposed to describe the momentum distribution
amplitude for the quark and antiquark in the $\vert q\bar {q}\rangle$
configuration. The asymptotic function was calculated  using perturbative 
QCD (pQCD) methods \cite{leb80,BroLep81,Bro82,EfrRad80,BerBroGol81}, 
and is the solution to the pQCD
evolution equation for very large $Q^2$ ($Q^2 \rightarrow \infty$):
\begin{equation}
   \phi_{as}(u) =\sqrt{3} u(1-u)
\;.\label{asy}\end{equation}
Using QCD sum rules, Chernyak and Zhitnitsky \cite{CheZhi84}
proposed  a function (CZ) that is 
expected to be correct for low Q$^2$:
\begin{equation}
   \phi_{cz}(u) =5\sqrt{3} u(1-u)(1-2u)^2
\;.\end{equation}
In recent experimental work \cite{e791} it was concluded that the 
asymptotic distribution amplitude describes the pion well 
for $Q^2 > 10\mbox{ (GeV/c)}^2$, see also Fig.~\ref{split}b.
\begin{figure} 
\begin{center}
   \resizebox{0.35\textwidth}{!}{
   \includegraphics{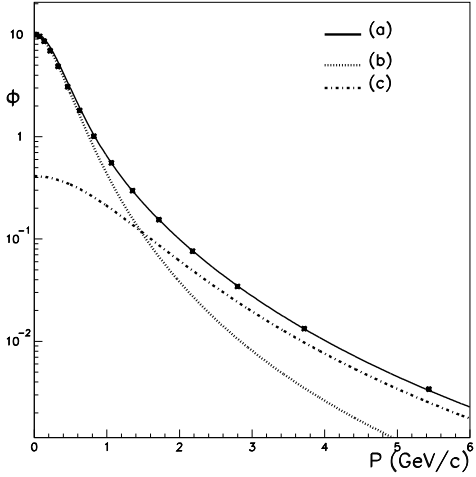}
   }\end{center}
\caption{Fits to the Singlet-Model wave function. 
   (a) is the fit to the full range, 
   (b) is the fit to the low range first term, and 
   (c) is the fit to the higher range, second term. 
   The resulting parameters are:
   $a = 0.7 $, $p_a =  0.515 \mbox{ GeV/c}$, 
   $b = 2.55$, $p_b =  1.58  \mbox{ GeV/c}$.
}\label{pawf} 
\end{figure}

Alternative approaches are based directly on the wave function, 
particularly for the non-perturbative low $Q^2$ region.
Jakob and Kroll \cite{JakKro93} proposed a Gaussian function: 
$\psi \sim e^{-\beta k_{\perp}^2}$.
Pauli \cite{Pau00} considers the $L_z=S_z=0$ component 
of the $u\bar d$ wave function in the Singlet Model: 
$\psi(u,\vec k_{\!\perp})\equiv 
\Psi_{u\bar d}(u,\vec k_{\!\perp};\uparrow\downarrow)$.
Then   
$H^{QCD}_{LC} \vert\psi_h\rangle  =  M^2_h \vert\psi_h\rangle$ 
translates to Eq.(8) of \cite{Pau00}: 
\begin{eqnarray} 
    &M^2&
    \psi(u,\vec k_{\!\perp}) =  
    \frac{ m^2 + \vec k_{\!\perp}^{\,2}}{u(1-u)} 
    \ \psi(u,\vec k_{\!\perp}) 
    -\frac{\alpha}{3\pi^2} \times
\\ 
    &\times&  \!\int\!\! 
    \frac{ du' d^2 \vec k_{\!\perp}'\ \psi(u',\vec k_{\!\perp}')}
    {\sqrt{ u(1-u) u'(1-u')}}
    \left(\frac{4 m^2}{Q ^2} + 
    \frac{2\mu^2}{\mu^2+Q^2}\right)
\;,\nonumber\end{eqnarray} 
for equal masses $m_1=m_2=m$, and for 
the mean Feynman four-momentum transfer $Q^2$ of the quarks.
Replacing the integration variable $u$ by $k_z$
according to
\begin{eqnarray}
   u = \frac{1}{2}
   \left[1+\frac{k_z}
   {\sqrt{m^2 + \vec k _{\!\perp}^{\,2} + k_z^2}}\right]
\;.\end{eqnarray}
Inversely, one expresses $k_z$ by $u$ with
\begin{eqnarray}
   k_z^2 = (m^2+\vec k_{\!\perp}^{\,2})
   \ \frac{\left(u-\frac{1}{2}\right)^2}{u(1-u)} 
\;.\label{eq:xkz}\end{eqnarray}
The substitution
allows to introduce the 3-vector 
$\vec p \equiv (\vec k_{\!\perp},k_z)$ and to
trans-scribe this integral equation into
\begin{eqnarray}
   & &\left[4m^2+4\vec p^{\,2}\right]
   \varphi(\vec p) - \frac{2\alpha}{3\pi^2}
   \int\frac{d^3\vec p\,'\ \varphi(\vec p\,')}
   {m\sqrt{A(p)A(p')}} 
   \times
\label{eq:10}\\ &\times&
   \left(\frac{4m^2}{Q^2} +
   \frac{2\mu^2}{\mu^2+Q^2} \right)
   = M^2\varphi(\vec p)
\;.\nonumber\end{eqnarray}
This equation was solved for $\varphi(\vec p)$ numerically 
\cite{Pau00}, in the non-relativistic approximation 
$A(p)=\sqrt{1+\vec p^2/m^2}\sim 1$.
In this approximation,
the mean four-momentum becomes $Q^2=(\vec p - \vec p\,')^{2}$,
and the light-cone wave function 
$\psi(u,\vec k_{\!\perp})$
is related to the reduced wave function 
$\varphi(k_z,\vec k _{\!\perp})$ 
by
\begin{eqnarray}
   \psi(u,\vec k _{\!\perp}) = 
   \ \frac{\varphi(k_z,\vec k _{\!\perp})}{\sqrt{u(1-u)}}
\;,\label{eq:psi}
\end{eqnarray} 
for details see \cite{Pau00}.
The parameters $\alpha = 0.69$, $m$ = 0.406 GeV, 
and $\mu$ = 1.33 GeV,
yield the correct mass-squared eigenvalue
$M^2=(140~\mathrm{MeV})^2$ for the pion and all other iso-scalar mesons.  

The isotropic numerical wave function $\varphi(\vec p)$ is
parametrized \cite{PauMuk01,Pau01b} 
for $p < 0.9 $ GeV/c as:
\begin{equation}
   \varphi(\vec p) = \frac{\mathcal{N}}
   {\displaystyle\left(p_a^2+\vec p^{\,2}\right)^{2}}
   ,\qquad p_a=0.515 \mbox{ GeV/c}
\;,\label{eq:ff} \end{equation}
that is, like a Coulomb wave function with an abnormally
large Bohr momentum: $p_a > m $.
 In Fig. \ref{pawf} we present a parametrization of the 
full range of the calculated wave function \cite{Pau01b} by a two-term 
Coulomb wave function. 
The first term is kept as in Eq. \ref{eq:ff} with
$p_a  =  0.515$ GeV/c. 
The mean momentum of the second term, $p_b$, 
and the two normalization constants $a,b$ are free parameters:
\begin{equation}
   \varphi (\vec p) = \frac{\displaystyle a}
   {\left(p_a^2 +{\vec p^{\,2}}\right)^2} + 
   \frac{b}{\left(p_b^2 + {\vec p^{\,2}}\right)^2}
\;.\label{eq:13}\end{equation} 
By extending the parametrization to the higher momentum range we can 
expect this function to be relevant to the measured $k_{\perp}$ 
distribution.

The light cone wave function (Eq. \ref{eq:psi})  
with the variable transform (Eq. \ref{eq:xkz}) becomes
\begin{eqnarray}
   \psi(u,\vec k _{\!\perp}) &=& 
   2\mathcal{N}\left[\frac
   {a\left[4u(1-u)\right]^{\frac{3}{2}}}
   {\left[\vec k _{\!\perp} ^{\,2} + K^2_a(u)\right]^2} +
   \frac
   {b\left[4u(1-u)\right]^{\frac{3}{2}}}
   {\left[\vec k _{\!\perp} ^{\,2} + K^2_b(u)\right]^2}\right]
\;,\nonumber\\ \mbox{with }
   K^2_a(u) &=& 4u(1-u) p_a^2 + (2u-1)^2 m^2
\;,\nonumber\\ 
   K^2_b(u) &=& 4u(1-u) p_b^2 + (2u-1)^2 m^2
\;.\label{eq:psiFinal}\end{eqnarray} 

The main experimental test of the wave function in the non-perturbative 
regime has traditionally been through measurements of the pion 
electromagnetic form factor \cite{SteSto97}. 
However, as shown by the authors, there is little 
sensitivity of the form factor to the distribution function. 
It is therefore desirable to develop alternative ways for these 
tests.

\section{Measurement of the Light-Cone Wave Function}
\label{sec:measwf}

The recent measurement of the pion light-cone wave function \cite{e791}
is based on the following concept: a high energy pion
dissociates diffractively on a heavy nuclear target. The first (valence)
Fock component dominates at large $Q^2$; the other terms are suppressed  
by powers of $1/Q^2$ for each additional parton, according to counting
rules \cite{SteSto97,BroFar73}. This is a coherent process in which the quark and 
antiquark break apart and hadronize into two jets. If in this 
fragmentation
process the quark momentum is transferred to the jet, measurement of the
jet momentum gives the quark (and antiquark) momentum. The proportionality 
of the differential cross section with respect to the jet momentum to the 
distribution amplitude (squared) was asserted \cite{FraMilStr93}:
\begin{equation}
   {d^3\sigma_N\over du\;dM^2_J\cdot d^2P_{N_t}} = 
   2.6\;{\rm GeV}^{-6}
   \left({{\rm GeV}\over \kappa_t}\right)^{8}\phi^2(u),
\end{equation}
and $u_{measured} = \frac {p_{jet1}} {p_{jet1}+p_{jet2}}$. 
$k_t$ is the measured transverse momentum of each jet. It is assumed that
$k_t(jet) \sim k_{\perp}(quark)$. This 
relation was also studied via Monte carlo simulations in order to 
verify the proportionality and take into account 
smearings in the fragmentation process and kinematic effects \cite{e791}.
From simple kinematics and assuming that the masses of the jets are small 
compared with the mass of the di-jets, the virtuality
and mass-squared of the di-jets are given by:
$Q^2 \sim  M_{DJ}^2 = \frac{k_t^2}{u(1 - u)}$ . 
The results of the measurement show that for $k_t > $1.5 GeV/c, which 
translates to $Q^2 > 10 {\rm (GeV/c)^2}$, there is good agreement 
between data and the asymptotic wave function. The conclusion was
that the pQCD approach that led to construction of the asymptotic wave 
function is reasonable for $Q^2 > 10 {\rm (GeV/c)^2}$. At lower values 
contributions from non-perturbative effects may become noticeable. In this 
work we focus our attention to this transition region.

\section{The $k_{t}$-Distribution}
\label{kperp}

The $k_t$ dependence of diffractive di-jets is another observable that can
show how well the various wave functions describe the data. As shown in
\cite{FraMilStr93} this dependence is expected to be: 
${d\sigma\over dk_t}  \sim  k_t^{-6}$ 
for one gluon exchange perturbative calculations. The 
experimental results are shown in Fig. \ref{split}, reproduced from Ref. 
\cite{e791}. In Fig. \ref{split}(a) the results were fitted by $k_t^n$ for 
$k_t > 1.25$ GeV
with n = $-9.2 \pm 0.5$. This slope is significantly larger than
expected. However, the region above $k_t \sim$ 1.8 GeV/c could be fitted
(Fig. \ref{split}(b)) with n = $-6.5 \pm 2.0$,
consistent with the predictions. This would support the evaluation of
the light-cone wave function at large $k_t$ as due to one gluon exchange,
as is the asymptotic wave function. 

The lower $k_t$-region can be considered as a transition from the 
perturbative to the non-perturbative regimes. 
The experimental results go 
down to $k_{t}  \sim $ 1 GeV/c, still large to be considered as a 
fully non-perturbative regime. In \cite{e791} this region was
fitted with the non-perturbative Gaussian
function: $\psi \sim e^{-\beta k_t^2}$ \cite{JakKro93}, resulting in
$\beta  =  1.78  \pm  0.1$.
Model-dependent values in the range of 0.9 - 4.0 were used \cite{JakKro93}.
This fit, although resulting in the parameter $\beta$ being consistent 
with theoretical expectations, is not very satisfactory. As seen in Fig. 
\ref{split}(b) the curved shape of the theoretical calculation is not 
observed in the data.

\begin{figure*}\sidecaption
\resizebox{0.70\textwidth}{!}{\includegraphics{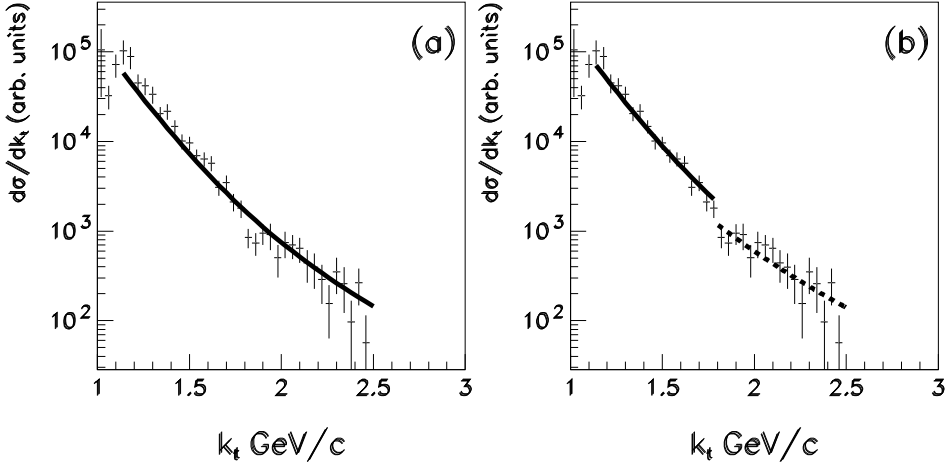}
}\caption{Comparison of the $k_t$ distribution of 
   acceptance corrected data with fits to
   cross section derived from 
   (a) two-terms wave function representing the full 
   range of the Singlet-Model wave function. 
   (b) Two-term wave function
   for low $k_t$, and a power law, 
   as expected from perturbative calculations, for high $k_t$.
}\label{coul} 
\end{figure*}

We now turn to comparison of the experimental data with the prediction of 
the  wave function based on the Singlet Model (section \ref{sec:int}). 
For this purpose we begin with the same expression used by the authors of 
\cite{e791},
\begin{equation}
   {d\sigma\over dk_t^2} \propto \left|\alpha_s(k_t^2)G(u,k_t^2)\right|^2
   \left|\frac{\partial^2}{\partial k_t^2}\psi (u,k_t)\right|^2
\;,\label{eq:16}\end{equation}
with $\alpha_s(k_t^2)G(u,k_t^2)  \sim  k_t^{\frac{1}{2}}$, 
which is based on factorization and 
which actually is a double differential cross section.   
Using it with the $\psi$ from
Eq. \ref{eq:psiFinal} gives
\begin{eqnarray}
   \frac{d^2\sigma}{du\ dk_t} = \mathcal{N}_1
   \ \frac {\left[4u(1-u)\right]^{3}}
   {k _t^{10}} \frac%
   {\displaystyle\left[1-         \frac12 \frac{K^2_a(u)}{k _t^2}\right]^2}
   {\displaystyle\left[1+\phantom{\frac12}\frac{K^2_a(u)}{k _t^2}\right]^8}
\;, \label{eq:d2sdudk}
\end{eqnarray} 
up to a constant $\mathcal{N}_1$, and up to similar terms with $K^2_b$.

The $k_{t}$-distribution is obtained from Eq.~\ref{eq:d2sdudk}
as a single differential cross section by integrating over $u$. 
But since $d^2\sigma/du\ dk_t$
is strongly peaked at $u=\frac12$, an exact treatment is
both complicated and unnecessary. Rather, it is approximated 
by removing the slowly varying terms from the integral,   
\begin{eqnarray}
   \frac{d\sigma}{dk_t} = \frac {\mathcal{N}_1}
   {k _t^{10}} \frac%
   {\displaystyle\left[1-         \frac12 \frac{\langle K^2_a\rangle}{k _t^2}\right]^2}
   {\displaystyle\left[1+\phantom{\frac12}\frac{\langle K^2_a\rangle}{k _t^2}\right]^8}
   \int_0^1\!\!\! du \left[u(1-u)\right]^3
\;,\label{eq:dsdk}
\end{eqnarray} 
and similarly for the $b$-terms. 
The average value $\langle K^2_a\rangle$ is evaluated
in two different approximations:
\begin{eqnarray}
   \langle K^2_a\rangle &=& \phantom{\frac89}p_a^2 
\label{eq:19}
\;,\\ \mbox{  and  }\qquad
   \langle K^2_a\rangle &=& \frac89 p_a^2 + \frac19 m^2
\;,\label{eq:20}
\end{eqnarray} 
and similarly for the $b$-terms.
The former is obtained by the peak value 
$\langle K^2_a\rangle=K^2_a(\frac12)$, 
the latter is obtained by the weighted average:
\begin{eqnarray}
   \langle u(1-u)\rangle &=& 
   \frac{\int_0^1\,du\;w(u)\ u(1-u)}{\int_0^1\,du\;w(u)} = \frac29 
\;,\label{eq:21}
\\ 
   \langle (2u-1)^2\rangle &=& 
   \frac{\int_0^1\,du\,w(u)\ (2u-1)^2}{\int_0^1\,du\;w(u)} = \frac19
\;,\label{eq:22}
\end{eqnarray} 
with the weight function $w(u)=\left[u(1-u)\right]^3$.
For the particular case $p_a=m$ both approximations coincide.

In Fig. \ref{coul}(a) we compare the prediction of Eq.~\ref{eq:19} 
with the data over the whole measured range. 
Only the total normalization constant $\mathcal{N}_1$ is a free parameter. 
The quality of the fit ($\chi^2$ = 1.7) improves significantly 
when we repeat the fit only in the low $k_t$ region ($\chi^2$ = 0.8). 
This is shown in Fig. \ref{coul}(b) with the higher $k_t$ 
range left as in Fig. \ref{split}(b). 

\section{\large \bf The $u$-Distributions}
\label{sec:4} 

Finally, we consider the $u$-dependence of the diffractive di-jets 
by integrating the double differental cross section (Eq. \ref{eq:d2sdudk}) 
over the transverse momentum $k_t$.
E791 \cite{e791} measures conditional $u$-distributions
\begin{eqnarray}
   \left.\frac{d\sigma}{du} \right\vert_{(k_l,k_u)}
   \;,\qquad\mbox{ with }\quad  
   k_l\leq k_t \leq  k_u
\;,\label{eq:23}\end{eqnarray}
for the momentum intervals  
$1.25\,\mathrm{GeV/c}\leq k_t \leq  1.5\,\mathrm{GeV/c}$ and 
$1.5\mbox{ GeV/c}\leq k_t \leq  2.5\mbox{ GeV/c}$. 
For comparing that with theory one should integrate
Eq. \ref{eq:d2sdudk}, and we do it in terms of the 
auxiliary function $G(u) = G(u;k_l)$: 
\begin{eqnarray}
   G(u;k_l) &\equiv& \int\!d^2k_t   
   \;\frac{d^2\sigma}{du\ dk_t}\;\theta(k_t^2-k_l^2)
\label{eq:24}\\
   &=& \pi\mathcal{N}_1 \left[u(1-u)\right]^3
   \int\limits_{k_l^2}^{\infty}\!dz\ z\;\frac%
   {\left[z-\frac12 K^2_a(u)\right]^2}
   {\left[z+\phantom{\frac12} K^2_a(u) \right]^8}
\;.\nonumber\end{eqnarray} 
The integration can be carried out analytically:
\begin{eqnarray}
   G(u;k_l) &=& \pi\mathcal{N}_1 \frac%
   {\left[u(1-u)\right]^3}
   {\left[k_l^2+ K^2_a(u) \right]^7} \times
\label{eq:25}\\
   &\times&\left[\displaystyle
   \frac{k_l^6}{4} - \frac{k_l^4}{20} K^2_a(u) +
   \frac{k_l^2}{40} K^4_a(u) + 
   \frac{1}{280} K^6_a(u) \right]
\;.\nonumber\end{eqnarray}
The conditional $u$-distributions become then
\begin{eqnarray}
   \left.\frac{d\sigma}{du} \right\vert_{(k_l,k_u)} = 
   G(u;k_l) - G(u;k_u)
\;.\label{eq:26}\end{eqnarray}
%
\begin{figure}  
\begin{center}\resizebox{0.48\textwidth}{!}{
   \includegraphics{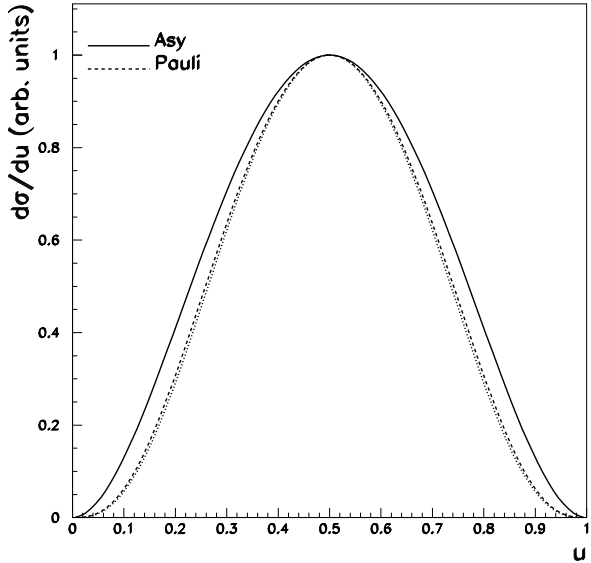}
}\end{center}
\caption{The asymptotic distribution $\left[u(1-u)\right]^2$ (solid
line) is compared with two $u$-distributions calculated for
$1.25\,\mathrm{GeV/c}\leq k_t \leq  1.5\,\mathrm{GeV/c}$, 
and $1.5\,\mathrm{GeV/c}\leq k_t \leq  2.5\,\mathrm{GeV/c}$ which  almost
coincide (dashed lines). All curves are normalized at $u$=0.5.
}\label{fig:4} 
\end{figure}
In Fig.~\ref{fig:4} we compare the predictions of Eq.~\ref{eq:26} 
to the asymptotic distribution $[u(1-u)]^2$ 
of \cite{leb80,BroLep81,Bro82,EfrRad80,BerBroGol81}. 
Only the overall normalization is used as free parameter. 
They are narrower than $[u(1-u)]^2$ mostly due to the
factor $[u(1-u)]^3$ in Eq.~\ref{eq:24}.

\section{\large \bf Discussion and conclusions}
\label{sec:5} 

The agreement of the calculated $k_{t}$-distribution 
with the data in the transition region of Fig.~\ref{coul} 
is interesting since the wave function is a non-perturbative
solution of a relatively simple model for the Light-Cone Hamiltonian. 
The model was developed for the bound state problem in
physical mesons \cite{Pau00},
where the low $k_{\!\perp}$-properties are important.
It is surprising that this model describes also the 
large $k_{\!\perp}$-properties in the tail of the wave function,
without that a re-fit of the parameters is necessary.

The non-relativistic approximation to Eq.~\ref{eq:10} predicts a
$u$-distribution quite similar to that predicted by the Asymptotic wave
function. This is a result of it being a solution of the Light-Cone
Hamiltonian for a $|q\bar{q}>$ system that conserves angular
momentum. Other non-perturbative wave functions such as that derived from
a harmonic oscillator potential do not have this property. The agreement
of the predicted $u$-distrubution with the E791 data \cite{e791} is not as
good as that of the Asymptotic wave function but this is not surprising as
these data were taken in a $k_t$ region that is in the $k_{\perp}$ tail of
this function.

The non-relativistic approximation to Eq.~\ref{eq:10} is a
technical simplification but not a compelling part of the model.
It can be relaxed in future work, without major difficulties.

In general, we can note 
that the $k_{\perp}$-distribution is a powerful tool for studying
wave functions in the perturbative regime and in the transition 
region between the perturbative and non-perturbative regimes.

The data of E791 were taken more than ten years ago. With realistic
light-cone wave functions now existing or coming up in the forseable
future, dedicated diffraction experiments, possibly extended to kaons and
other hadrons, should be very valuable.

\end{document}